# Estimates of daily ground-level NO$_2$ concentrations in China based on big data and machine learning approaches

Xinyu Dou,[1] Cuijuan Liao,[1] Hengqi Wang,[1] Ying Huang,[2] Ying Tu,[1] Xiaomeng Huang,[1] Yiran Peng,[1] Biqing Zhu,[1] Jianguang Tan,[1] Zhu Deng,[1] Nana Wu,[1] Taochun Sun, [1]Piyu Ke,[1] Zhu Liu[1]*

**Affiliations**
1. Department of Earth System Science, Tsinghua University, Beijing 100084, China.
2. Department of Hydraulic Engineering, Tsinghua University, Beijing 100084, China.

*Correspondence: Zhu Liu; e-mail: zhuliu@tsinghua.edu.cn

**ABSTRACT:**
Nitrogen dioxide (NO$_2$) is one of the most important atmospheric pollutants. However, current ground-level NO$_2$ concentration data are lack of either high-resolution coverage or full coverage national wide, due to the poor quality of source data and the computing power of the models. To our knowledge, this study is the first to estimate the ground-level NO$_2$ concentration in China with national coverage as well as relatively high spatiotemporal resolution (0.25 degree; daily intervals) over the newest past 6 years (2013-2018). We advanced a Random Forest model integrated K-means (RF-K) for the estimates with multi-source parameters. Besides meteorological parameters, satellite retrievals parameters, we also, for the first time, introduce socio-economic parameters to assess the impact by human activities. The results show that: (1) the RF-K model we developed shows better prediction performance than other models, with cross-validation R$^2$ = 0.64 (MAPE = 34.78%). (2) The annual average concentration of NO$_2$ in China showed a weak increasing trend (0.17±0.46 μg/$m^3$ $yr^{-1}$). μg/m$^3$ yr$^{-1}$ While in the economic zones such as Beijing-Tianjin-Hebei region, Yangtze River Delta, and Pearl River Delta, the NO$_2$ concentration there even decreased or remained unchanged, especially in spring. Our dataset has verified that pollutant controlling targets have been achieved in these areas. With mapping daily nationwide ground-level NO$_2$ concentrations, this study provides timely data with high quality for air quality management for China. We provide a universal model framework to quickly generate a timely national atmospheric pollutants concentration map with a high spatial-temporal resolution, based on improved machine learning methods.

**KEYWORDS:**
Ground-level NO$_2$ concentration, China, Random Forest model, Multi-source big data, Socio-economic parameters, K-means

## 1. INTRODUCTION

Since the Industrial Revolution, a great deal of energy, such as coal and oil, has been used in human activities with the rapid development of the economy and society. The combustion process of energy has caused a large number of atmospheric pollutants, and global environmental pollution has become increasingly serious. $NO_2$ is one of the important atmospheric pollutants monitored by countries around the world. It is also an important precursor of acid rain, tropospheric ozone, and atmospheric aerosol, which is harmful to ecology, animal, and human health[1]. China is the largest developing country in the world and has been suffering from serious $NO_2$ pollution due to the large emissions from fuel combustion by automobiles and power generation[2,3]. Regionally, population densely areas such as Beijing-Tianjin-Hebei (BTH) region, Yangtze River Delta (YRD), and Pearl River Delta (PRD) are faced with the most severe $NO_2$ pollution problems[4-6]. According to the World Health Organization's safety standards for $NO_2$ pollution—an average annual concentration below 40 $μg/m^3$—China still needs to reduce the level of $NO_2$ pollution as the concentration in many regions far exceeds this standard (National Bureau of Statistics of China, http://www.stats.gov.cn/).

Data on ground-level $NO_2$ concentrations are critical for emission mitigation and policy making. Since 2013, China has set up more than 1,000 state-managed sites to monitor the concentration of $NO_2$ and other atmospheric pollutants in the environment. However, it is still quite difficult to accurately quantify the ground-level $NO_2$ concentration nationwide. The numbers of ground observation stations are still far from adequate. In addition, ground observation stations are unevenly distributed. Moreover, there is a high degree of uncertainty in assessing exposure levels of $NO_2$ through site matching, especially for remote areas in western China with few stations. Therefore, national coverage of ground-level accurate prediction of the spatial and temporal distribution of $NO_2$ across the country is essential to population exposure assessment.

To have a full "panoramic view" of $NO_2$ concentrations across the country, physical mechanism models and empirical statistical models (such as simple linear regression models, land-use regression models, geographically weighted regression models) have been applied extensively[7-12], but they suffer from some potential limitations. The physical mechanism models rely on the atmospheric physical and chemical transmission mode coupled with satellite-retrieved vertical column density (VCD) of $NO_2$ to estimate the ground-level concentration. However, the structure of this model is complicated, and its performance is heavily depending on the setting of key parameters of the atmospheric physical and chemical reaction process and relying on relatively huge computing resources[13-15]. As for the empirical statistical models, they integrate meteorological and other auxiliary factors to simulate the relationship of satellite-retrieved VCD of $NO_2$ and the ground-level $NO_2$ concentration. So far, this approach has advanced from simple linear regression models[16,17][16,17] to an advanced statistical analysis that integrates satellite retrievals and multi-source geographic covariates, such as land-use analysis[18,19], geographically weighted regression[20,21], etc. However, big challenges remain in simulating $NO_2$ for the whole country based on statistical methods. First, the spatial resolution of the $NO_2$ dataset generated by traditional statistical models is relatively low, which limits its application in small- and medium-scale areas (e.g., urban). Second, due to a large amount of calculation and the lag of some data, the models cannot be updated in real-time. Third, the formation mechanism of ground-level $NO_2$ is complex, with many influencing factors and large variation. Traditional statistical models are not sufficient to fully explain the complex nonlinear and high-order interaction relationships between $NO_2$ and influencing factors[17-23].

Compared with traditional statistical methods, machine learning approaches (such as

Random Forest models, Light Gradient Boosting Machine models, and Neural Network models) generally show higher prediction accuracy[24,25]. They develop complex model structures to capture relationships that would otherwise be too complex to specify in parametric models[22]. Meanwhile, compared with physical mechanism models, the simulation method based on machine learning approaches has a lower cost of establishing models, and there are lower requirements for data and hardware performance. In addition, machine learning approaches have fast training speed and efficient processing of large datasets and can be well applied in air pollution forecasting based on nearly-real-time big data. Now, machine learning models are increasingly used to extract patterns and insights from the Earth system data, as well as in the prediction of multiple atmospheric pollutants, such as fine particulate matter ($PM_{2.5}$, $PM_1$) and $O_3$[26-33]. However, they have rarely been applied for $NO_2$ data, especially in China[34-36]. In the rare studies on $NO_2$ in China, researchers only took satellite retrievals and geographic covariates as predictors, neglecting the important role of social-economic predictors in the ground-level $NO_2$ concentration, though they strongly affect anthropogenic $NO_2$ emissions[36,37].

In this study, we introduced a new machine learning approach to estimate the ground-level $NO_2$ concentration in China and consequently obtain high-resolution maps of the national ground-level $NO_2$ concentration. As shown in Figure 1, first, we collected and standardized multi-source big data to obtain a dataset with uniform spatial-temporal resolution (0.25 degrees, daily resolution). Despite of the traditional parameters using for estimation, such as satellite retrievals parameters and meteorological parameters, we also added the social-economic parameters with the consideration of impact by human activities. Second, we constructed a Random Forest model integrated K-means to estimate the relationship between the aforementioned parameters and the ground truth ($NO_2$ concentrations from monitoring sites), which shows better performance than the traditional linear regression model as well as the ordinary Random Forest model. Finally, a newest and complete spatial-temporal distribution of the national ground-level $NO_2$ concentration dataset from 2013 to 2018 in China was built in this study, with a temporal resolution of daily and spatial resolution of 0.25 degrees. This dataset not only presents the spatiotemporal variation of the ground-level $NO_2$ concentrations nationwide, but also make up for the lack of areas due to poorer data quality. With this dataset, we can deepen our understanding on the ground-level $NO_2$ distribution in China nationwide, which will give us confidence in our ability to develop strategies for either air quality management or epidemiological management in China.

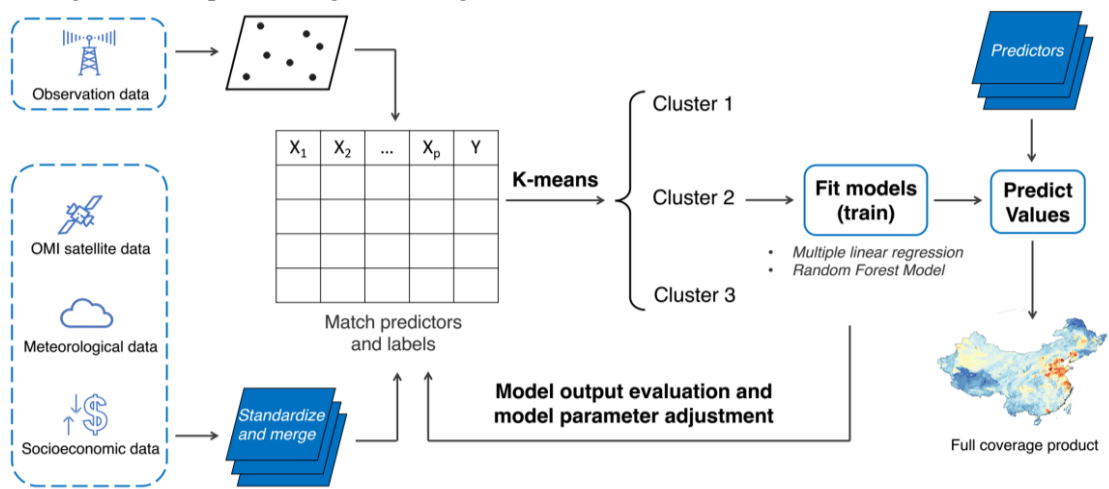

**Figure 1. The flow diagram of this study.**

## 2. MATERIALS AND METHODS

### 2.1 Multi-Source Big Data

In this study, three categories of data (including satellite retrievals data, social-economic data, and meteorological data) a total of 14 variables (Table S1) were used as predictors, and the Ground-level $NO_2$ observations were regarded as model labels to construct a supervised machine learning model. After data cleaning and spatial-temporal unification, 680131 pieces of data were obtained for model training. Detailed information about the data collection and processing is given in the **Supplemental Material**.

### 2.2 Models

In this study, a traditional statistical model (Multiple Linear Regression model) and a machine learning model (Random Forest model) are used to fit the $NO_2$ observations with the predictors mentioned above. The entire model programming will be based on sklearn package in Python. For the detailed algorithm of the Multiple Linear Regression Model and Random forest Model, please refer to **Supplemental Methods**. And these two models are constructed based on the data after k-means clustering, so we call them MLR-K (Multiple Linear Regression model integrated K-means) and RF-K (Random Forest model integrated K-means) respectively.

Different clustering results will be tested for comparison to find the Random Forest model with the best performance. The parameters of the Random Forest model do not significantly affect the accuracy of the final model, which makes it easier to train the data for each category. After automatic parameter tuning by cross validation, the final RF-K consists of 300 regression trees on every cluster, which is based on 300 bootstrap samples randomly selected from the training data. Nearly a third of the predictive variables were randomly selected to build each tree to reduce the correlation between the trees. And the MLR-K will be compared as a reference.

## 3. RESULTS

We constructed a 0.25°*0.25°daily ground-level $NO_2$ concentrations dataset based on big data and a Random Forest model integrated K-means(RF-K).

Before running the model, we dealt with the predictors' correlation analysis first. For detailed result analysis, please see **Supplemental Results**. After experiments, we found that the K-means clustering can effectively use the information of the input predictors, greatly improve the accuracy of the prediction model. Please see **Supplemental Results** for detailed analysis of clustering results via K-means. In the following, we integrated K-means with Multiple Linear Regression model and Random Forest model to improve their performances.

## 3.1 Model performance assessment

This study used the following Taylor diagram to explicitly characterize the model performance. This diagram, invented by Karl E. Taylor in 1994 (published in 2001) facilitates the comparative assessment of different models[38]. It is used to quantify the degree of correspondence between the modeled and observed behavior in terms of three statistics: the Pearson correlation coefficient, the root-mean-square error (RMSE) error, and the standard deviation. Two models based on K-means, each represented by a different color on the diagram, were compared, and the distance between each model and the point labeled "observation" is a measure of how realistically each model depicts observations. For each model, three statistics were plotted: the Pearson correlation coefficient (gauging similarity in pattern between the simulated and observed fields) is related to the azimuthal angle (black contours); the centered RMS error in the simulated field is proportional to the distance from the point on the x-axis identified as "observations" (red contours); the standard deviation of the simulated pattern is proportional to the radial distance from the origin (blue contours).

It can be seen from the Taylor diagram that in the performance comparison of the two models, the traditional Multiple Linear Regression models integrated K-means(MLR-K) all show obvious weakness on both the general model and four seasons sub-models (see Figure 2 and Figure S8). The MLR-K model has a lower coefficient of determination $R^2$ (0.3804) and a higher root-mean-square error, which indicates that the predicted results of the MLR-K model do not fit well with the observations. Also, the standard deviation of the MLR-K model is significantly lower than the standard deviation of the observations (18.97 $\mu g/m^3$). It indicates that the prediction results of the MLR-K model cannot effectively simulate the variability of $NO_2$ concentration between different regions. In comparison, the Random Forest model integrated K-means (RF-K) shows a higher fitting degree, which is reflected in the higher coefficient of determination $R^2$ (0.6419) and lower root-mean-square error. Therefore, this study generates an improved daily ground-level $NO_2$ concentration dataset of China with a high spatial resolution based on the RF-K model.

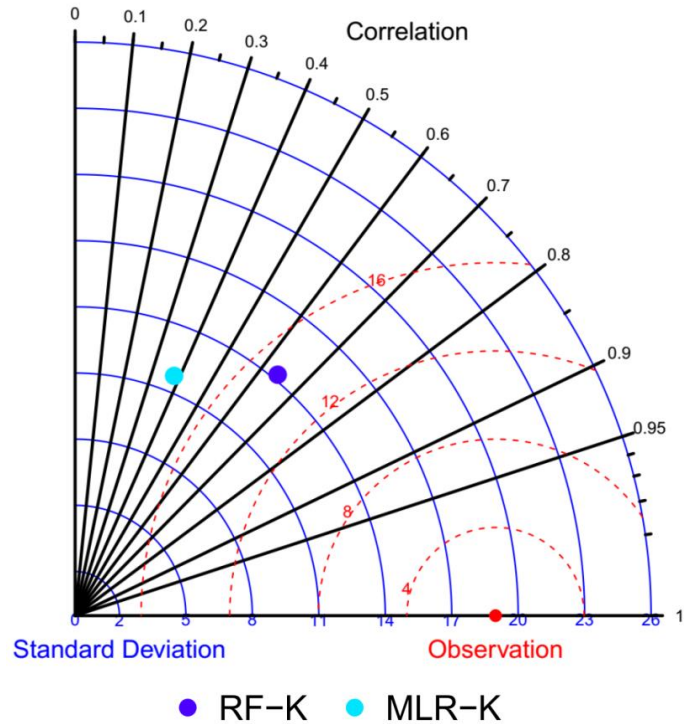

**Figure 2. Taylor diagram of the Multiple Linear Regression model integrated K-means(MLR-K) and the Random Forest model integrated K-means(RF-K).** Note: The radial distance from the dot represents the standard deviation of the model. The closer the standard deviation of the model is to the standard deviation of the observation, the better the fitting ability of the model. The dashed red semicircle with the observation (red dot) as the center represents the root-mean-square-error, which represents the distance between the observation and the model. The closer the model point is to the observation (red dot), the closer distance it is. The correlation coefficient is determined by the azimuth position of the model. When the model simulation result is more consistent with the observation, the closer the model point is to the observation point on the x-axis, it means that the model has a higher correlation with the observation.

As shown in Figure 3, RF-K model also has better fitting accuracy. The predicted results of the RF-K model are in good agreement with the observations (Figure 3). The overall fitting $R^2$ of the RF-K model is 0.64, and the RMSE value is 11.36 μg/$m^3$(Figure 3). In contrast, the MLR-K model has a poor-fitting effect, the fitting $R^2$ is only 0.38, and the RMSE value increases to 14.94 μg/$m^3$(Figure 3). The MLR-K model shows significant underestimation in high concentration areas, suggesting that the linear relationship of predictors alone could not accurately simulate a high level of ground-level $NO_2$ concentrations. It is not sensitive to seasonal changes (Figure S9). As Figure 3 shown, compared with the MLR-K model, the RF-K model based on machine learning can better simulate and reveal the complex relationship between ground-level $NO_2$ concentration and impact factors, which makes the overall fitting accuracy of the RF-K model greatly better than that of the MLR-K model. In addition, 10-fold cross validation was adopted to test the model stability. After ten data transformations, the score of the model has little change, which proves that the model does not overfit due to special data(Figure S10).

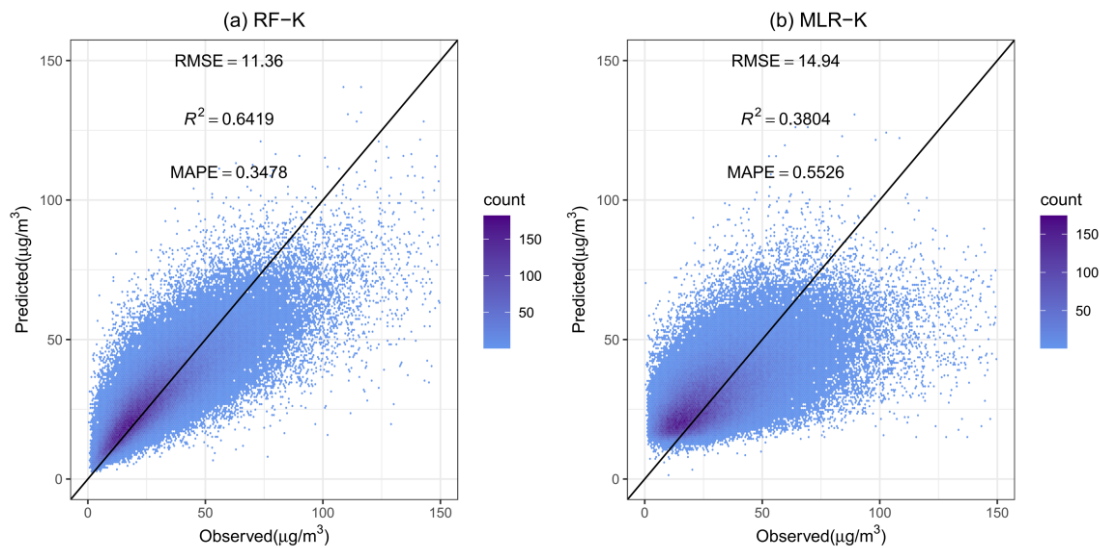

**Figure 3. Performance of these two models in predicting daily NO$_2$ concentrations for China.**
The solid black lines represent 1:1.

A reasonable K-means model will help us understand the regional predictors and improve the accuracy of the model. K-means is mainly grouped into a category with a large difference in NO$_2$ concentration gradient based on predictors, which can be reflected in geographical location distribution (Figure S7). The contribution of emission capacity of different regions to local NO$_2$ concentration varies greatly, and this information is difficult to be observed in a large database. When we apply K-means, the use of predictors to distinguish the clusters with large differences can be conducive to the observation of areas with special conditions. For model training, a large amount of data is needed to provide more learning information. At the same time, too large a data set will also reduce the simulation capacity of the model for individual quantities due to the diversity of its variables. To maintain a balance between the amount of data and the accuracy of the model, we explored the reasonable number of categories of K-means. As shown in Table S2, when K-means is not conducted (the number of clusters is 1), it can be seen that the model's determination coefficient R$^2$ is about 0.6400 and RMSE is about 11.3889 μg/$m^3$. With the number of clusters ranging from 2 to 8, we found that the model had the best performance when the number of clusters is 3, with R$^2$ reaching 0.6419 and RMSE decreasing to 11.3568 μg/$m^3$ (Table S2). The accuracy (1-MAPE) of the final RF-K model is about 65.3%. Due to the huge difference in the data of four seasons, here we had an experiment on the data of seasonal division, the accuracy of the model is further improved, from the original accuracy rate (1-MAPE) of 65.3% to 68.2% (Table S2). It suggests that the RF-K model can be established by seasons when sufficient data are available in future applications.

### 3.2 Predictors importance

We also analyzed which predictors play a leading role in the RF-K construction, as shown in Figure 4. It is shown that OMI-NO$_2$ is the most important predictor in the model, with relative importance values ranging from 0.2019 to 0.2486(Figure 4). This is mainly because the OMI-NO2 data reflects the NO$_2$ vertical column density in the atmosphere, which is closely related to the ground-level $NO_2$ concentration.

The importance distribution diagram of predictors implies that the model is based on satellite retrievals data, with socio-economic data as the main cofactors. Satellite retrievals are based on grid-based instantaneous measurements, while $NO_2$ observations are based on daily average observations on monitoring sites. As expected, this spatial-temporal mismatch tends to reduce the strength of the correlation between these two variables. Due to spatial mismatch, the accuracy of predicting ground-level $NO_2$ concentration only from satellite retrievals will be greatly reduced. While the results of predictors' importance also show that in addition to OMI satellite data, the combination of social-economic and meteorological predictors can effectively reduce this mismatch. Socio-economic and meteorological factors could adjust the estimated distribution of ground-level $NO_2$ concentration to make it different from the $NO_2$ vertical column density in the atmosphere reflected by satellite retrievals.

The newly incorporated social-economic predictors (nightlight data, artificial impervious area data) in this study showed great importance in predicting ground-level $NO_2$ concentration (Figure 4). Closely following the OMI satellite data, the importance of the two social-economic predictors ranked as second and third most important. Nightlight data and artificial impervious area data reflect the significant interference effect of surface human activities on ground-level $NO_2$ concentration. The importance of social-economic predictors provides a scientific evidence that by incorporating human activities, the accuracy of subsequent ground-level air pollutants prediction models can be greatly improved.

The importance value of the meteorological condition predictors for different clusters ranges from 0.0452 to 0.0988(Figure 4), which is consistent with the fact that meteorological predictors greatly affected the environmental fate of $NO_2$. Wind speed, as the most important meteorological predictor, has a negative impact on $NO_2$, which indicates that advection is an important process to remove $NO_2$ pollution.

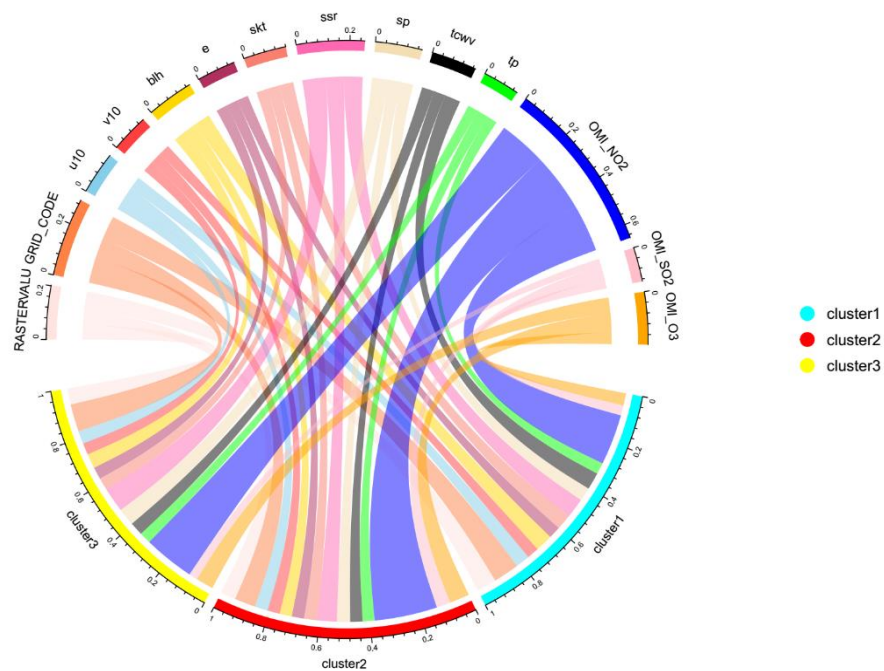

**Figure 4. The relative importance of the predictor variables in the RF-K Model (3 clusters).**

### 3.3 Estimates of ground-level NO$_2$ concentration

**3.3.1 Spatial distribution**

We averaged and plotted the annual average ground-level NO$_2$ concentration maps during 2014 and 2017 (Table S3 and Figure 5).

Generally, the annual ground-level NO$_2$ concentration across China keeps at a high level (33.46±6.31 μg/$m^3$). High concentrations of NO$_2$ are mainly distributed in Beijing-Tianjin-Hebei region, Yangtze River Delta, and Pearl River Delta. This is mainly due to intense human activities and serious pollutant emissions. In contrast, the areas of Southwest, Northeast, and South China with less human activities or better climatic conditions have much lower NO$_2$ concentrations. It is worth noting that the unique local topographical features in Xinjiang and Inner Mongolia affect the accumulation of NO$_2$. The high NO$_2$ concentration in these regions is inconsistent with reality. In the future, we plan to reduce this error by optimizing the screening of predictors.

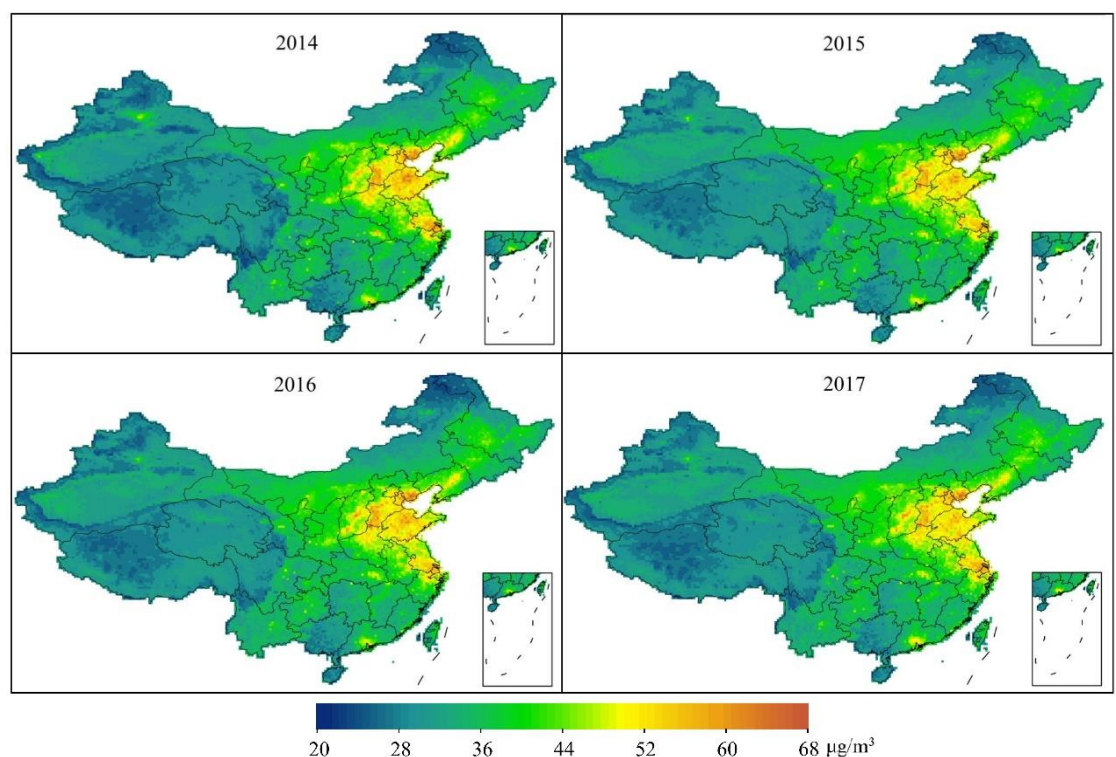

**Figure 5. Spatial distributions of annual mean NO$_2$ concentrations from 2014 to 2017 across China.**

Ground-level NO$_2$ concentration varies on a seasonal scale, as shown in Table S3 and Figure 6. Overall, there is no significant difference (33.30~33.82 μg/$m^3$) in the average seasonal changes of NO$_2$ in China (Table S3). This is likely due to the vast territory of China. When averaging across the whole country, the differences in different periods will be smoothed out. Whereas, there are prominent variations for local regions, especially in the North China

Plain (Figure 6). In spring and winter, NO$_2$ pollution is serious among the main economic zones of China, such as Beijing-Tianjin-Hebei region, Yangtze River Delta, and Pearl River Delta, show a high NO$_2$ concentration. It can be explained by the burning of coal and fossil fuels caused by human activities. In contrast, NO$_2$ concentration is lower in summer and autumn because the frequent rainfall and the conducive weather conditions promote the diffusion of pollutants and reduce air pollution.

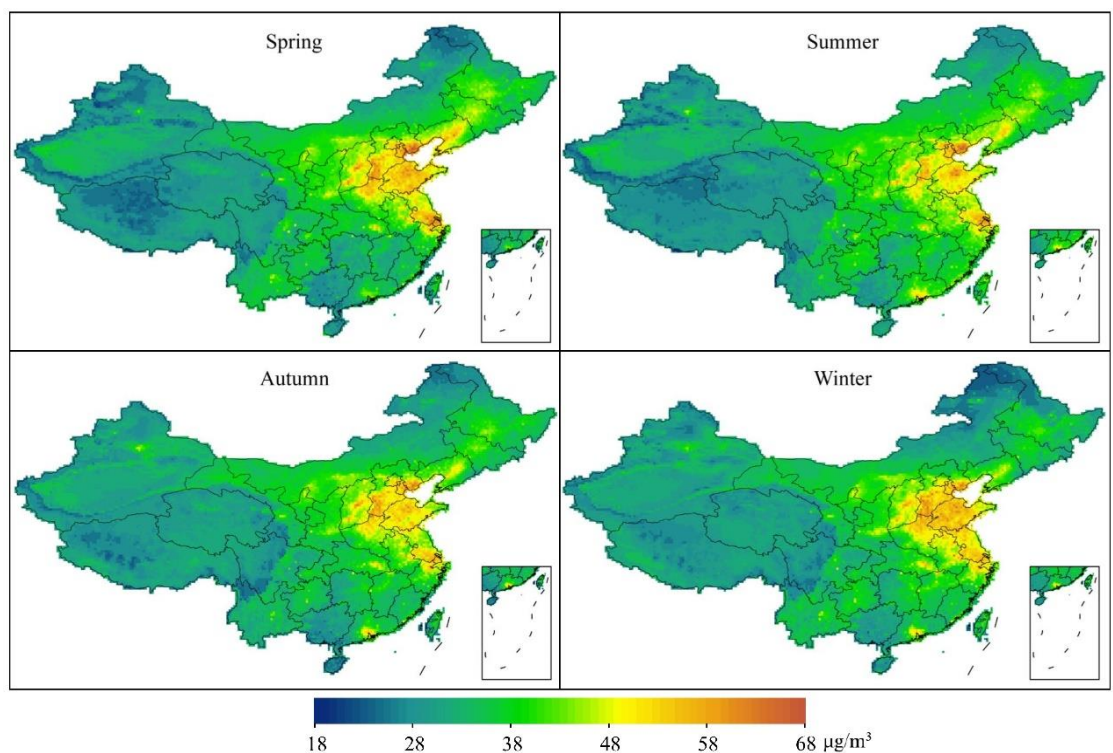

**Figure 6. Spatial distributions of seasonal average NO$_2$ concentrations during 2013 and 2017 across China.** Note: The warmer the color, the greater the concentration, and vice versa.

**3.3.2 Temporal Trend**

To explore the inter-annual and inter-seasonal changes of the ground-level NO$_2$ concentration in various regions of China, we have drawn the spatial distribution map of the NO$_2$ change trend from 2014 to 2017, as shown in Figure 7.

In general, the average concentration of ground-level NO$_2$ in China has shown a slightly increasing trend (0.17±0.46 μg/$m^3$ $yr^{-1}$), with the most obvious increase in summer (0.23±0.58 μg/$m^3$ $yr^{-1}$), less significant change in spring (0.09±0.60 μg/$m^3$ $yr^{-1}$) (Table S4). However, if we focus on specific regions (Figure 7), the increasing trend mainly appears in the central and western regions such as Inner Mongolia and Xinjiang, which have less observation and relatively poor simulation results. The emergence of the increasing trend may be related to our simulation deviation. However, for some economically more developed areas such as Beijing-Tianjin-Hebei region, Yangtze River Delta, and Pearl River Delta, the NO$_2$ concentration shows decrease or invariability, especially in spring. The above results suggest that some environmental protection policies in these areas may have achieved initial results in recent years.

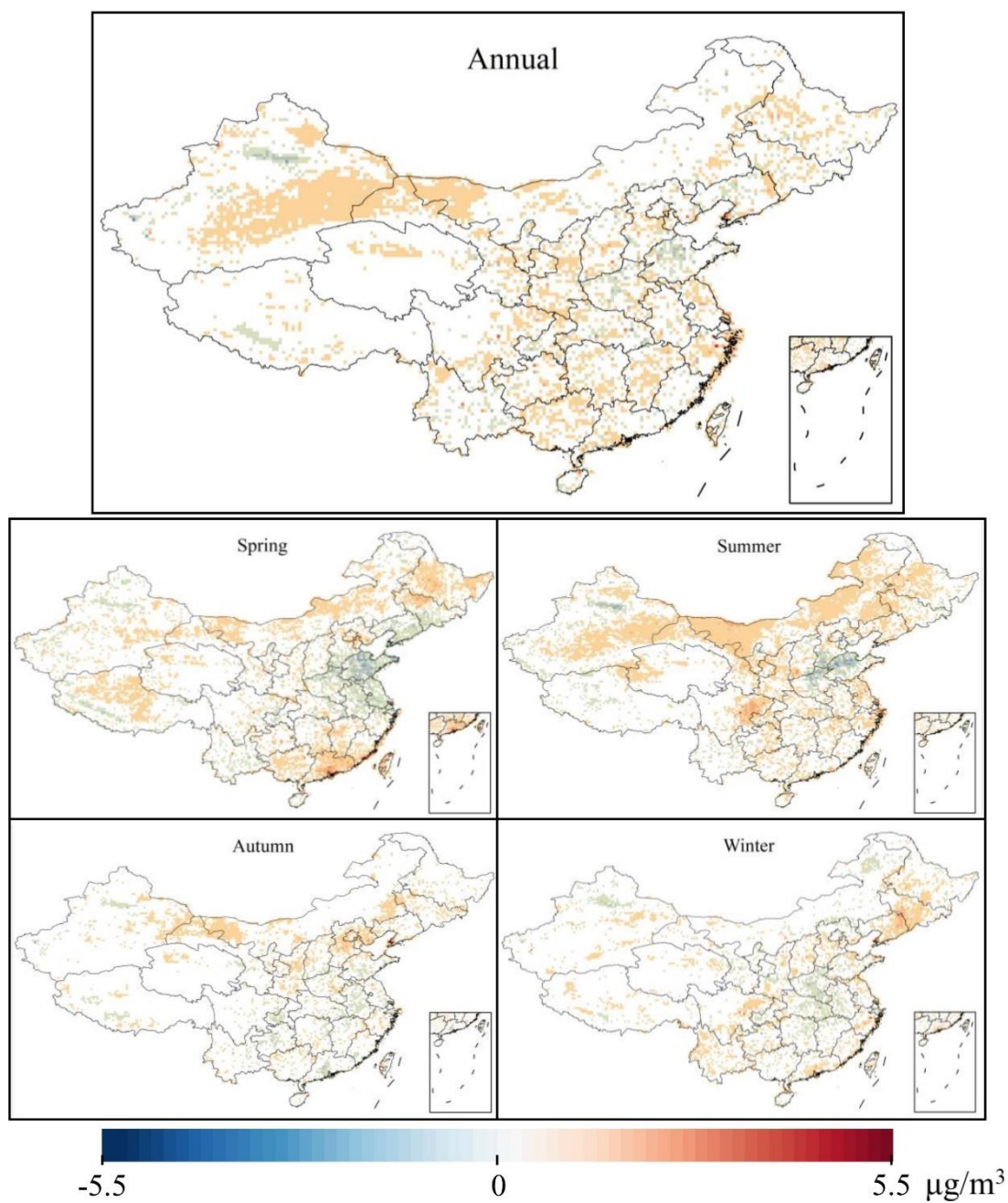

**Figure 7. Spatial distribution for annual (top) and seasonal (bottom) trends of ground-level NO₂ in China from 2013 to 2018.** Note: The warmer the color, the greater the growth rate, and vice versa.

### 3.4 Uncertainty analysis

Our 10-fold cross-validation results (detail see Figure S10) show that randomly selected data does not affect the accuracy of the RF-K model, which reflects the robustness of the model. The model has a high degree of adaptability to new data, which is conducive to real-time updating of training models and prediction. Even if there are many dynamic variables and interference factors on the daily scale, we can already greatly realize the prediction with the daily resolution. If the model is applied to monthly, seasonal, and annual scale studies, the accuracy of the model will be further improved, and the uncertainty will be further reduced.

Obviously, since our training data comes from the satellite retrievals and reanalysis database, there is a certain degree of uncertainty when gridding these data. While the fluctuation of this part of the data can be completely regarded as the noise of the model to train the stability of the model. It will not be particularly sensitive to fluctuations in subsequent predictive factors to produce biased prediction results.

From the final results (Figure 3), the goodness of fit of the Random Forest model integrated K-means we finally adopted can reach more than 0.64, and the average error only accounts for about 34.78% of the original data (MAPE=34.78%), which means that the error fluctuations are concentrated on the upper and lower 34.78% of the observation data. According to the observation data, the $NO_2$ concentration on the same day can fluctuate up to about 50% due to the temperature difference between day and night, the peak period of commuting to and from get off work, etc. So the error between our model's prediction results and observation data is acceptable. Also, it can be seen from Figure 2 that the standard deviation of the model prediction results is about 14.25 μg/$m^3$, while the observation data is 18.97 μg/$m^3$, which indicates that the model prediction results have sufficient precision. Considering that observations data are far from adequate due to the limitation of observation time and observation space, the prediction results of our model can be able to fill the data gap because it is a full coverage estimation.

## 4. DISCUSSION

In general, the dataset of this study is consistent with most of existing dataset for China with different spatial resolutions, and show a reasonable spatial-temporal distribution and changing trends from 2013 to 2018(see Table S4). Particularly, our dataset is the newest, and the model performance is higher than previous similar research results (Table 1), which is very important for providing high-quality basic data for air quality management. In terms of model performance, the estimated daily data $R^2$ of the RF-STK model proposed in the past research is 0.62, and the RMSE is 13.3 μg/$m^3$ [36], which is lower than our model performance($R^2$=0.64, RMSE=11.4 μg/$m^3$). Some models reduce the time resolution (monthly) to obtain an $R^2$ of 0.84 and an RMSE of 6.33 μg/$m^3$ [39]. And some other traditional models, such as Land Use Regression Model (LUR) got $R^2$=0.57[40]. The RF-K model developed for China in this study can outperform most previous similar models indicated by improved performance assessment indicators (of almost all categories). This is mainly because: (1) Our model considers the important role of socioeconomic factors; (2) We perform reasonable interpolation from available satellite data to regions that cannot be fully covered from real measurements. The daily nationwide ground-level $NO_2$ concentrations map generated in this study provides timely and detailed data with high quality, which is extremely valuable for real-time air pollution research in China.

**Table 1. Performance comparison between this study and the previous models in predicting ground-level $NO_2$ concentration.**

| Reference | Model | Study Period | Validation | Metric | Daily Resolution | Spatial Resolution |
|---|---|---|---|---|---|---|
| This study | Random Forest integrated K-means clustering(RF-K) | 2013-2018 | 10-fold site-based cross validations | $R^2 = 0.64$ RMSE = 11.4μg/$m^3$ | Daily | 0.25° |
| Zhan Y et al.[36] | Random-forest-spatio-temporal-kriging(RF-STK) | 2013-2016 | 10-fold site-based cross validations | $R^2 = 0.62$ RMSE = 13.3μg/$m^3$ | Daily | 0.1° |
| You J et al.[39] | Random Forest(RF) | 2015-2016 | 10-fold site-based cross validations | $R^2 = 0.84$ RMSE = 6.33μg/$m^3$ | Monthly | 13km×13km |
| Shi Y et al.[40] | Linear regression (Land Use Regression, LUR) | 2016 | Fitting | $R^2 = 0.571$ | Anuual | 0.5km×0.5km |

The predicted results show that the national level of $NO_2$ still showed a weak increasing trend during 2013-2018(0.17±0.46 μg/$m^3$ $yr^{-1}$), while the major economic zones such as Beijing-Tianjin-Hebei region, Yangtze River Delta and Pearl River Delta decreased or remained unchanged in concentration. The improvement of air quality in these three economic zones has benefited from the government's high attention, perfect air pollution prevention and control laws and policies, detailed and highly enforceable pollution prevention and control measures, and strong supervision and supervision mechanisms. Although the air quality in these three economic zones has been significantly improved, our results show that China's nationwide $NO_2$ pollution prevention and control still faces huge challenges. China's highly polluting energy structure heavily relying on coal, gasoline, and diesel is an important reason for its high $NO_2$ concentration. Therefore, it is urgent for China to develop new energy to replace coal fuels and to take other measures to further improve the energy structure. In addition, the number of motor vehicles in China is increasing rapidly, but vehicle standards and oil products cannot keep up with the international development level, which is also an important reason for the huge increase in $NO_2$ emissions. Promoting the upgrading of oil products and developing new energy vehicles to reduce the $NO_2$ emission in the transportation sector are vital to improving air quality for China.

Despite the advances in the model and modeled results, this study still has a few points that should be addressed in the future. Firstly, the existing model could be developed into a model with temporal and spatial migration capabilities, so that the parameters of the model can be automatically adjusted according to the newly added training data. Such a model will be able to realize the effective prediction of future prediction of the ground-level $NO_2$ concentration. Secondly, although the accuracy of our RF-K model developed in this study has reached a high level, it can still be further improved by incorporating more predictors and with better model designs. In the follow-up study, our model with temporal and spatial migration capabilities can make it possible to continue to update daily concentrations at the nation level, which can make it possible to assess persistent changes in ground-level $NO_2$ concentrations. High resolution and timely-updated ground-level $NO_2$ concentration data can contribute to comprehensively monitoring local air quality changes and accurately identifying heavily polluted areas for key remediation, so as to shorten the response time of policy adjustments.

## DESCRIPTION OF SUPPLEMENTAL INFORMATION

Supplemental Information includes four material sections, three methods sections, two results sections, eleven figures, four tables and supplemental references.

## DECLARATION OF INTERESTS

The authors declare no competing interests.

## ACKNOWLEDGMENTS

Authors acknowledge the National Natural Science Foundation of China (grant 71874097 and 41921005), Beijing Natural Science Foundation(JQ19032), and the Qiu Shi Science & Technologies Foundation.
We thank Professor Qiang Zhang of Tsinghua University for providing ground-level $NO_2$ observations data and valuable suggestions.

## AUTHOR CONTRIBUTIONS

Xinyu Dou: Writing-Original Draft, Methodology, Visualization, Writing-Reviewing and Editing. Cuijuan Liao: Methodology, Software, Editing. Hengqi Wang: Methodology, Software, Editing. Ying Huang: Resources. Ying Tu: Resources. Xiaomeng Huang: Reviewing and Editing. Qiang Zhang: Reviewing and Editing. Yiran Peng: Reviewing and Editing. Biqing Zhu: Reviewing. Jianguang Tan: Reviewing. Zhu Deng: Reviewing. Nana Wu: Resources. Taochun Sun: Editing. Piyu Ke: Visualization. Zhu Liu: Conceptualization, Supervision, funding acquisition.

## FIGURE TITLES AND LEGENDS

**Figure 1. The flow diagram of this study.**

**Figure 2. Taylor diagram of the Multiple Linear Regression model integrated K-means(MLR-K) and the Random Forest model integrated K-means(RF-K).** Note: The radial distance from the dot represents the standard deviation of the model. The closer the standard deviation of the model is to the standard deviation of the observation, the better the fitting ability of the model. The dashed red semicircle with the observation (red dot) as the center represents the root-mean-square-error, which represents the distance between the observation and the model. The closer the model point is to the observation (red dot), the closer distance it is. The correlation coefficient is determined by the azimuth position of the model. When the model simulation result is more consistent with the observation, the closer the model point is to the observation point on the x-axis, it means that the model has a higher correlation with the observation.

**Figure 3. Performance of these two models in predicting daily NO$_2$ concentrations for China.** The solid black lines represent 1:1.

**Figure 4. The relative importance of the predictor variables in the RF-K Model (3 clusters).** Note: Refer to Table S1 for the detailed descriptions of the variables.

**Figure 5. Spatial distributions of annual mean NO$_2$ concentrations from 2014 to 2017 across China.**

**Figure 6. Spatial distributions of seasonal average NO$_2$ concentrations during 2013 and 2017 across China.** Note: The warmer the color, the greater the concentration, and vice versa.

**Figure 7. Spatial distribution for annual (top) and seasonal (bottom) trends of ground-level NO$_2$ in China from 2013 to 2018.** Note: The warmer the color, the greater the growth rate, and vice versa.

# TABLE TITLES AND LEGENDS

**Table 1. Performance comparison between this study and the previous models in predicting ground-level NO$_2$ concentration.**

| Reference | Model | Study Period | Validation | Metric | Daily Resolution | Spatial Resolution |
|---|---|---|---|---|---|---|
| This study | Random Forest integrated K-means clustering(RF-K) | 2013-2018 | 10-fold site-based cross validations | $R^2 = 0.64$ RMSE = 11.4μg/$m^3$ | Daily | 0.25° |
| Zhan Y et al.[36] | Random-forest-spatio-temporal-kriging(RF-STK) | 2013-2016 | 10-fold site-based cross validations | $R^2 = 0.62$ RMSE = 13.3μg/$m^3$ | Daily | 0.1° |
| You J et al.[39] | Random Forest(RF) | 2015-2016 | 10-fold site-based cross validations | $R^2 = 0.84$ RMSE = 6.33μg/$m^3$ | Monthly | 13km×13km |
| Shi Y et al.[40] | Linear regression (Land Use Regression, LUR) | 2016 | Fitting | $R^2 = 0.571$ | Anuual | 0.5km×0.5km |